\documentclass[aps,reprint,twocolumn,nofootinbib]{revtex4-1}
\usepackage{graphicx,amsfonts,amsmath,amssymb}
\usepackage{bm}
\usepackage{url}
\makeatletter
\g@addto@macro{\UrlBreaks}{\UrlOrds}
\makeatother

\begin{document}
\title{Cyclotron Beam Extraction by Acceleration}
\date{\today}
\author{C. Baumgarten}
\affiliation{Paul Scherrer Institute, Switzerland}
\email{christian.baumgarten@psi.ch}

\def\begeq{\begin{equation}}
\def\endeq{\end{equation}}
\def\begary{\begeq\begin{array}}
\def\endary{\end{array}\endeq}
\def\bmtx{\left(\begin{array}}
\def\emtx{\end{array}\right)}
\newcommand{\myarray}[1]{\begin{equation}\begin{split}#1\end{split}\end{equation}}
\def\eps{\varepsilon}
\def\g{\gamma}
\def\y{\gamma}
\def\w{\omega}
\def\W{\Omega}
\def\s{\sigma}
\def\beginfig{\onecolumngrid}
\def\endfig{\onecolumngrid}
\def\Exp#1{\exp\left(#1\right)}
\def\Log#1{\ln\left(#1\right)}
\def\Sinh#1{\sinh\left(#1\right)}
\def\Sin#1{\sin\left(#1\right)}
\def\Tanh#1{\tanh\left(#1\right)}
\def\Tan#1{\tan\left(#1\right)}
\def\Cos#1{\cos\left(#1\right)}
\def\Cosh#1{\cosh\left(#1\right)}
\def\pt{p_\theta}

\begin{abstract}
  One of the decisive issues in the design and operation of cyclotrons
  is the choice of the beam extraction method. Typical methods are extraction
  by electrostatic extractors and by stripping. The former method requires
  DC high voltage electrodes which are notorious for high-voltage breakdowns. 
  The latter method requires beams of atomic or molecular ions which are 
  notorious for rest gas and Lorentz stripping.
  Here we discuss the conditions to be met such that a beam will leave the
  magnetic field of an isochronous cyclotron purely by fast acceleration.   
\end{abstract}

\keywords{Cyclotrons, Particle Accelerators, High Intensity Accelerators, Beam
Extraction, Accelerator Driven Systems}
\maketitle

\section{Introduction}

The extraction method is a decisive choice for the design of
isochronous cyclotrons~\cite{Richardson,Craddock,Stammbach1992,Onishchenko2008,
Calabretta2016,Seidel2021,Seidel2018,Kleeven2006}. Specifically in 
case of high intensity beams, extraction losses have to be minimized 
in order to avoid activation of the machine components~\cite{Blaser1963}. 
The extraction methods used most frequently are stripping 
extraction~\cite{RICKEY1962,SCHWABE1996,Djurovic2001,Mackenzie1984,MEC} and 
extraction by electrostatic deflectors~\cite{Martin1963,Gunder1963}. 
Both methods have their advantages and issues.

Stripping extraction, i.e. the removal of electrons by the passage
of an ion beam through thin stripper foils, requires the
acceleration of ions that are not fully stripped yet, for instance of
$H^-$-ions or $H_2^+$-molecules. But since not fully ionized atoms
(or molecules) have a considerably larger scattering cross-section
with the rest gas molecules, some losses in the course of acceleration are 
unavoidable. Besides rest gas stripping also Lorentz stripping may result
in losses~\cite{Hiskes1961,Stinson1969,Keating1995,LStrip,Calvo2021}. 
Therefore stripping extraction requires an excellent vacuum and the 
Lorentz stripping effect limits both, 
the maximal beam energy and the maximal magnetic field.
Low-loss extraction by electrostatic deflectors, on the other hand,
is notorious for beam interruptions by high voltage breakdowns
and requires well-separated turns in order to place the extraction
septum between turns.

Despite these issues, isochronous cyclotrons are attractive for the
production of high intensity CW beams, due to their superior energy
efficiency~\cite{Grillb2017}, their small footprint, and their 
relatively low cost. 
Therefore the use of cylotrons for the production of high intensity 
beams has been suggested in various projects, typically aiming for 
several $mA$ of beam current at energies between
$600\,\rm{MeV}$ and $1\,\rm{GeV}$, for instance for ADS systems,
neutron production and also neutrino physics~\cite{STAMMBACH1996,Alenitsky2011,WINKLEHNER2018,Yang2016}. 
A cyclotron facility that often serves as a proof-of-principle-machine 
for these objectives is the high intensity proton accelerator 
(HIPA) facility of the Paul Scherrer Institute (PSI) in Switzerland, 
which provides up to $2.4\,\rm{mA}$ of protons at $590\,\rm{MeV}$~\cite{hipa_acc}.

In 1981 Werner Joho formulated ``Joho's $N^3$ law'', which states that
the possible beam intensity, for the same losses, increases with the 
inverse of the third power of the number of turns~\cite{Joho1981}.
This law has been verified with astonishing accuracy at the PSI
Ring cyclotron~\cite{hipa_acc}. Hence we can reasonably assume that any 
high intensity proton machine will aim for the lowest possible number
of turns, that is, for the highest possible acceleration voltage, mostly
in order to provide the highest possible turn separation at extraction.
The incredible increase of beam current from the PSI Ring machine
has been achieved by the insertion of a flat-topping cavity and by the
reduction of the number of turns, from originally more than
$300$ to now $\approx 180$~\cite{hipa_acc} with an upgrade of the
rf cavities and amplifier chains. However, the reduction of the turn
number has a side effect which has not received much attention, namely
that the total phase shift of the beam by the fringe field near the
outer radius is significantly reduced.

In 1995, Yves Jongen proposed the so-called ``self-extraction'',
a method to design cyclotrons such that the beam would leave the field
without stripper or electrostatic extractor~\cite{Jongen1995}. 
A cyclotron build by IBA provided the proof-of-principle~\cite{Kleeven2001}
for this extraction method. However, a theoretical account of the conditions 
that have to be met to allow for beam escape has, to the knowledge of the
author, not been provided so far.

Here we report about the fact that the design of high energy high intensity
separate sector cyclotrons of the PSI type meets the main requirements 
for auto-escape of the beam without electrostatic deflectors. 

\section{The Cyclotron Bending Limit}

The main aim of the reduction of the turn number in the Ring cyclotron
is the increase of the turn separation so that the septum of the
electrostatic extractor can be placed between cleanly separated turns.
This is required not only in order to avoid an overheating of the
septum, but also to minimize beam loss and activation of the septum
and subsequent beamline elements.
Isochronous cyclotrons are efficient because they operate at constant 
magnetic field and constant rf-frequency. This allows for narrow-band 
rf structures with high Q-factors. Furthermore the beam
passes the same rf-structure multiple times which also increases the
efficiency of the acceleration. However, this method requires that the
circulation frequency of the beam stays in sync with the frequency
of the rf system. The average magnetic field must then increase radially
with the relativistic $\y$-factor and since the velocity is approximately
given by $v=\w\,R$, the (average) field must approximately follow
\begeq
B\propto{1\over\sqrt{1-(\w\,R/c)^2}}
\endeq
Therefore the isochronism of a cyclotron can not be sustained
in the fringe field due to the radial decrease of the magnetic field.
Hence the phase between beam and accelerating rf will shift in the 
course of extraction and the bunches will get more and more out of
sync with the accelerating rf. 

Let the energy gain per turn $dE/dn$ be given by
\begeq
dE/dn=q\,V_{rf}\,\cos{(\phi)}=\Delta E_{max}\,\cos{(\phi)}
\label{eq_egain}
\endeq
where $\phi$ is the phase of the beam (relative to the rf-phase) and
$V_{rf}$ is the maximum accelerating voltage per turn.
If the beam is not extracted fast enough, the phase $\phi$ will be shifted
beyond $90^\circ$ and the beam will loose instead of gain energy when
passing the next rf cavity.

Hence there is another important consequence of the reduction of turn number
in isochronous cyclotrons: The maximum energy that the beam is able
to reach depends crucially of the phase shift in the fringe field.
On the other hand, it is evident that any finite field $B(R)$ can
only hold a circulating beam up to a finite momentum.

The relation between momentum $p$, radius $R$ and magnetic field $B$
is given by
\begeq
p=q\,B\,R
\endeq
so that
\begeq
{dp\over dR}=q\,B\,(1+{R\over B}\,{dB\over dR})=q\,B\,(1+k)
\label{eq_dpdR}
\endeq
The factor $k={R\over B}\,{dB\over dR}$ is the field index\footnote{
The usual convention is $k=-{R\over B}\,{dB\over dR}$, but
the cyclotron literature mostly uses the positive sign convention. 
} 
The maximum momentum is given by ${dp\over dR}=0$ which corresponds
to a field index of $k=-1$. Beyond the point, where the radial field 
gradient in the fringe field region is steeper than $-B/R$, the beam
can not stay within the machine. Hence there is a maximum radius
\begeq
R_{max}=-{B(R_{max})\over \left.{dB\over dR}\right\vert_{R_{max}}}
\endeq
which corresponds to the maximum momentum
\begeq
p_{max}=q\,R_{max}\,B(R_{max})=-{q\,B(R_{max})^2\over \left.{dB\over dR}\right\vert_{R_{max}}}
\endeq
Let us call this momentum and the corresponding energy the escape
momentum and escape energy.

Hence there are two maximum values for the radius, firstly the 
radius where the phase shift reaches $90^\circ$ and secondly the 
radius of the maximum momentum. The decisive question is therefore, 
which of these radii is larger. This depends on two factors, firstly the
exact shape of the fringe field and secondly, the accelerating
voltage $V_{rf}$. The latter is in the general case a function of radius
$V=V(R)$ as well, but since this dependency is usually small (over the 
region of interest, i.e. the extraction), we shall neglect it in the following.

The reduction of the turn number has, as mentioned before, the main purpose
to increase the turn separation.
Since energy, radius and momentum have mutual bijective relationships, the
radius gain per turn ${dR\over dn}$ in a cyclotron is, in sectorless
approximation, given by
\begeq
  {dR\over dn}={dE\over dn}\,({dE\over dp}\,{dp\over dR})^{-1}\,.
\endeq
From Hamilton's equation of motion it is known that ${dE\over dp}=v$, so that
one obtains with Eq.~\ref{eq_egain} and Eq.~\ref{eq_dpdR}:
\begeq
  {dR\over dn}={V_{rf}\,\cos{(\phi)}\over v\,B\,(1+k)}={q\,V_{rf}\,\cos{(\phi)}\,R\over v\,p\,(1+k)}\,,
\endeq
which can be reformulated as
\begeq
  {dR\over dn}={q\,V_{rf}\,\cos{(\phi)}\,R\,\y\over E\,(\y+1)\,(1+k)}\,.
\label{eq_dRdn}
\endeq
Both, energy and radius vary only a little over the region of interest.
The dominating factors are therefore $\cos{(\phi)}$ and $1+k$.
The question is then, whether the radius for $\phi>=90^\circ$ or the
radius for $k=-1$ is smaller. When the phase $\phi$ approaches $90^\circ$
{\it before} $k$ approches $-1$, then the turn separation will typically
decrease to zero and then become negative, so that the $E-\phi$-loop closes.
However, if the phase $\phi$ stays well below $90^\circ$ when $k$ approches $-1$, then
the beam will escape the field simply because the momentum exceeds the
bending limit. 

\section{Estimation of the Acc. Voltage Required to Reach the Escape Energy}

If $\theta$ is the azimuthal angle and $\phi_{rf}$ the phase of the rf,
then the phase shift per time ${d\phi\over dt}$ in an isochronous cyclotron
can be written as
\myarray{
  {d\phi\over dt}&={d\phi_{rf}\over dt}-N_h\,{d\theta\over dt}\\
                 &=\w_{rf}-N_h\,\w_{rev}
}  
where $\w_{rev}$ is the actual revolution frequency of the bunch and
$N_h$ is the so-called harmonic number, that is the number
of rf-cycles per revolution of the bunch.
The number of revolutions per time is
\begeq
{dn\over dt}={1\over T_{rev}}
\endeq
so that
\myarray{
  {d\phi\over dn}&=T_{rev}\,(\w_{rf}-N_h\,\w_{rev})\\
  &=2\,\pi\,({\w_{rf}\over\w_{rev}} -N_h)\\
\label{eq_phaseshift0}
}
where $T_{rev}=2\,\pi\,R/v$ is the time required per revolution.
The relation between the revolution frequency of a particle with
charge $q$ and mass $m$ in the magnetic field $B(R)$ can be written as
\begeq
\w_{rev}={q\over m\,\y}\,B(R)\,.
\endeq
In perfectly isochronous machines the field is given by
\begeq
B_{iso}={B_0\over\sqrt{1-({R\,\w_0\over c})^2}}
\endeq
where $\w_0=q/m\,B_0=\w_{rf}/N_h$ is the ``nominal'' revolution frequency. 
Often the cyclotron radius $R_\infty=c/\w_0$ is used to write this as
\begeq
B_{iso}={B_0\over\sqrt{1-(R/R_\infty)^2}}=B_0\,\y_R
\endeq
where $\y_R$ is understood purely as a function of the radius. 

\begin{figure}[t]
\parbox{8cm}{
\includegraphics[width=80mm]{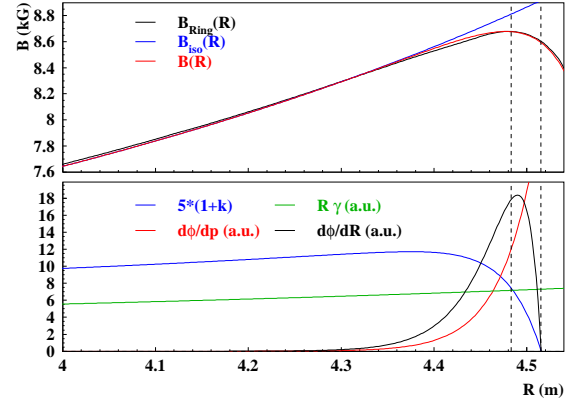}
}\caption{
    Top: Average fringe field $B_{Ring}$ of the PSI Ring Cyclotron,
    corresponding isochronous field $B_{iso}=5.45\,\rm{kG}\,\cdot\y_R$ and $B(R)$ according
    to Eqn.~\ref{eq_Bapprox} with $g=36\,\mathrm{mm}$
    and $R_h=4.63\,\rm{m}$ and $R_\infty=5.7\,\rm{m}$. 
    Bottom: The corresponding phase shift
    per momentum increase, phase shift per radius increase and 
    function $1+k$ (scaled by a factor of $5$).
    The positions of $k=0$ and $k=-1$ are indicated by dashed lines.
    The radius $R$, the factor $\y$ and even a factor $R\,\y_R$ (shown)
    vary only weakly over the fringe field region.
\label{fig_fringe}}
\end{figure}

Let us assume that the the fringe field can be approximated by an Enge 
type function~\cite{Enge1964,Enge1967} of the simplest form so that 
the real (azimuthally averaged) magnetic field $B(R)$ is given by
\begeq
B(R)=B_{iso}\,f(R)
\label{eq_Bapprox}
\endeq
with
\begeq
f(R)=(1+\exp{({R-R_h\over g})})^{-1}
\endeq
where $g$ is approximately half of the pole air gap~\footnote{
The exact value depends on the details of the iron geometry.} 
and $R_h$ is the radius 
for which the field is half of the isochronous field, i.e. $f(R_h)=1/2$.
The radial derivative of $f(R)$ is then
\begeq
{df\over dR}=-\frac{1}{g}\,f\,(1-f)\,.
\label{eq_dfdR}
\endeq
Fig.~\ref{fig_fringe} shows how this approximation compares to the
(azimuthal) average field of the PSI Ring cyclotron.
For our purposes the agreement is -- within the region of interest --
reasonable, even though the parameters obtained from this ``fit'' do
not agree very well with the real Ring cyclotron.
The field index is then given by
\myarray{
  k&={R\over B}\,{dB\over dR}\\
    &={R\over B}\,B_0\,\left({d\y_R\over dR}\,f+\y_R\,{df\over dR}\right)\\
    &=\y_R^2-1-(1-f)\,{R\over g}\\
  \label{eq_k}  
}
The revolution frequency can then be expressed by
\begeq
\w_{rev}={q\over m\,\y}\,B_0\,\y_R\,f(R)={\y_R\over\y}\,\w_0\,f(R)\,,
\endeq
where $\y$ is the usual relativistic factor, that is, a function of
velocity:
\begeq
\y={1\over\sqrt{1-v^2/c^2}}={1\over\sqrt{1-\beta^2}}\,.
\endeq
In the sectorless approximation the velocity is given by
\begeq
v=\w_{rev}\,R=R\,{\y_R\over\y}\,\w_0\,f\,,
\label{eq_velo}
\endeq
so that $\y$ is given by
\begeq
\y={1\over\sqrt{1-({R\over R_\infty}\,{\y_R\over \y}\,f)^2}}\,,
\endeq
This can be used to find the radial dependency of $\y$:
\begeq
\y=\sqrt{1+(\y_R^2-1)\,f^2}\,.
\label{eq_gamma_of_R}
\endeq
If the cyclotron is isochronous up to the fringe region, then
$\w_{rf}=N_h\,\w_0$ and therefore Eq.~\ref{eq_phaseshift0} yields
\begeq
  {d\phi\over dn}=2\,\pi\,N_h\,({\y\over\y_R\,f}-1)\,.
\label{eq_phshift}
\endeq
The phase shift per energy gain ${d\phi\over dE}$ can be expressed, using Eq.~\ref{eq_egain}, by
\myarray{
  {d\phi\over dE}&={d\phi\over dn}\,{1\over dE/dn}\\
                 &={2\,\pi\,N_h\over q\,V_{rf}\,\cos{(\phi)}}\,({\y\over\y_R\,f}-1)\\
}
and per momentum gain by
\begeq
  {d\phi\over dp}={d\phi\over dE}\,{dE\over dp}={d\phi\over dE}\,v\,.
\endeq
The velocity $v={dE\over dp}$ can be replaced by the use of Eq.~\ref{eq_velo} so that:
\begeq
  {d\phi\over dp}={2\,\pi\,N_h\,\w_0\over q\,V_{rf}\,\cos{(\phi)}}\,(1-{\y_R\,f\over\y})\,R\,.
\label{eq_dphdp}
\endeq
In order to obtain the phase shift as a function of radius, one may use
Eq.~\ref{eq_dpdR} to obtain
\myarray{
  {d\phi\over dR}&={d\phi\over dp}\,{dp\over dR}\\
&={2\,\pi\,N_h\,\w_0\over q\,V_{rf}\,\cos{(\phi)}}\,(1-{\y_R\,f\over\y})\,R\,q\,B(R)\,(1+k)\\
&={2\,\pi\,N_h\,m\,c^2\over q\,V_{rf}\,\cos{(\phi)}}\,{R\over R_\infty^2}\,(1-{\y_R\,f\over\y})\,\y_R\,f\,(1+k)\\
\label{eq_dphdR}
}
Introducing the abbreviation
\begeq
A={2\,\pi\,N_h\,m\,c^2\over q\,V_{rf}}
\label{eq_Adef}
\endeq
together with Eq.~\ref{eq_k} and Eq.~\ref{eq_gamma_of_R} one obtains:
\myarray{
{d\phi\over dR}&={A\over\cos{(\phi)}}\,{R\over R_\infty^2}\,(1-{\y_R\,f\over\sqrt{1+(\y_R^2-1)\,f^2}})\,\times\\
&\times\,\y_R\,f\,(\y_R^2-(1-f)\,R/g)\,.
}
and hence
\myarray{
d\sin{(\phi)}&=A\,{R\,\y_R\over R_\infty^2}\,(1-{\y_R\,f\over\sqrt{1+(\y_R^2-1)\,f^2}})\,\times\\
&\times\,f\,(\y_R^2-(1-f)\,R/g)\,dR\,.
}
As shown in Fig.~\ref{fig_fringe}, the phase $\phi$ and the term $1+k$ vary fast over the fringe field region, 
while the relative change of $R$ and $\y_R$ are small. Hence it is a reasonable approximation
to keep the latter constant in the integration. This means that we fix $R\approx R_x$ and 
$\y_R\approx\y_x$, where the subscript ``x'' indicates that these are the values at extraction.

From Eq.~\ref{eq_dfdR} one obtains
\begeq
dR=g\,{df\over f\,(f-1)}\,.
\endeq
One may express the phase factor by its Taylor series (using $f$ as variable, located at $f=1$):
\begeq
(1-{\y_R\,f\over\sqrt{1+(\y_R^2-1)\,f^2}})={1-f\over\y_R^2}+\frac{3}{2}\,{\y_R^2-1\over\y_R^4}\,(1-f)^2+\dots
\endeq
and since $1-f$ is small, one may use the first term only. Then the integrand
simplifies to:
\begeq
  d\sin{(\phi)}=-A\,g\,{R_x\over R_\infty^2\,\y_x}\,(\y_x^2+(f-1)\,R_x/g)\,df\,.
\endeq
\begin{figure}[t]
\parbox{8cm}{
\includegraphics[width=80mm]{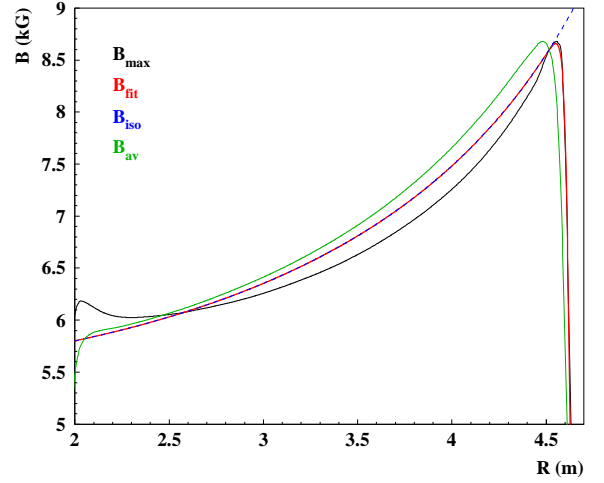}
}\caption[Field Shape of Ring Cyclotron]{
  Shape of the average ($B_{av}$, green) and maximum ($B_{max}$, black) magnetic field of
  the PSI Ring cyclotron\footnote{Compare to Fig. 6 in Ref.~\cite{Willax1969}.}. $B_{max}$ is 
  not isochronous, but the fall-off of the fringe field is steeper than that 
  of $B_{av}$. The shape $B_{fit}$ (red) provides the best fit around the maximum. The fit 
  is best with $g=15\,\rm{mm}$ (instead of $g=36\,\rm{mm}$ as derived from $B_{av}$). 
\label{fig_fringe2}}
\end{figure}
The integration over the fringe field region starts with $f\approx 1$ and ends
where $k=-1$, which corresponds (see Eq.~\ref{eq_k}) to $f=1-\y_R^2\,\frac{g}{R}$, so that the leading
term after integration is:
\myarray{
  \sin{(\phi_f)}-\sin{(\phi_i)}&\approx A\,{g^2\over
    R_\infty^2}\,{\y_x^3\over 2}\\
}
where $\phi_i$ is the initial phase (prior to extraction) and $\phi_f$ is
the final phase. If the initial phase is zero ($\sin{(\phi_i)}\approx 0$),
the condition $\phi\le 90^\circ$ yields $\sin{(\phi_f)}\le 1$, then the accelerating
voltage which suffices to reach the escape energy, is finally (skipping the
subscript $x$):
\begeq
q\,V_{rf}\ge{\pi\,N_h\,m\,c^2\,\y^3\,g^2\over R_\infty^2}
\label{eq_Vrfmin}
\endeq
With $\beta\approx R/R_\infty$ and $\Delta E_{max}=q\,V_{rf}$ this can also be
written as
\begeq
{\Delta E_{max}\over E}\ge\pi\,N_h\,\y\,(\y+1)\,{g^2\over R^2}\,,
\endeq
where $E=m\,c^2\,(\y-1)$ and $R$ are the (kinetic) extraction energy and radius. Since
the number of turns is approximately $N_t\approx E/\Delta E_{max}$, one finds
\begeq
N_t\le {1\over \pi\,N_h\,\y\,(\y+1)}\,{R^2\over g^2}
\label{eq_maxturns}
\endeq
Hence it is mostly the squared ratio of extraction radius and
pole gap at extraction which determines the maximal number of
turns or the minimal energy gain, respectively.
The compact superconducting cyclotron COMET~\cite{Schillo,Geisler2007},
which provides the $250\,\rm{MeV}$ proton beam for the proton
therapy facility Proscan at PSI, has a half-gap of $22\,\rm{mm}$,
an extraction radius of $820\,\rm{mm}$ and a harmonic number of $N_h=2$. 
According to Eq.~\ref{eq_maxturns}, escape extraction is then
possible for a maximum turn number of about $85$. This compares to
an actual turn number of about $650$, i.e. eight times as much.
Hence the beam would be able to escape without extractor under these
conditions, if the half-gap would be (reduced to) less than $8\,\rm{mm}$.

However, for the PSI Ring cyclotron, Eq.~\ref{eq_Vrfmin} results,
using the parameters obtained from the azimuthal average field
(i.e. $g\approx 36\,\rm{mm}$), in a minimal voltage of 
\begeq
V_{rf}\ge 2.95\,\rm{MV}
\endeq
The numerical integration of Eq.~\ref{eq_dphdR} for the same parameters
yields a voltage of $V_{rf}\ge 2.9\,\rm{MV}$ as required to reach the 
escape energy, which is in reasonable agreement with our approximation.
This voltage is close to the average voltage actually used today in the 
PSI Ring cyclotron, which is about $2.6\,\rm{MV}$. 

But the azimuthal average of the field is not a good approximation
for separate sector cyclotrons. A realistic value of the half-gap at
extraction is about $21\,\rm{mm}$. As shown in Fig.~\ref{fig_fringe2},
the fall-off of the {\it maximal} field values is much steeper than that
of the azimuthal average. Specifically the parameter $g$ is less than
half of the value obtained from the azimuthal average (and closer
to the real half-gap)\footnote{The most accurate determination of the
  field shape as ``seen by the beam'' would derive from the average
  field along the scalloping closed orbits. But stable closed orbits
  do not exist beyond $k=-1$.
}.
Hence the required voltage might well be a factor of $4$ smaller, 
i.e. as low as $V_{rf}\ge 650\,\rm{kV}$ or even less.

\begin{figure}[t]
\parbox{8cm}{
\includegraphics[width=80mm]{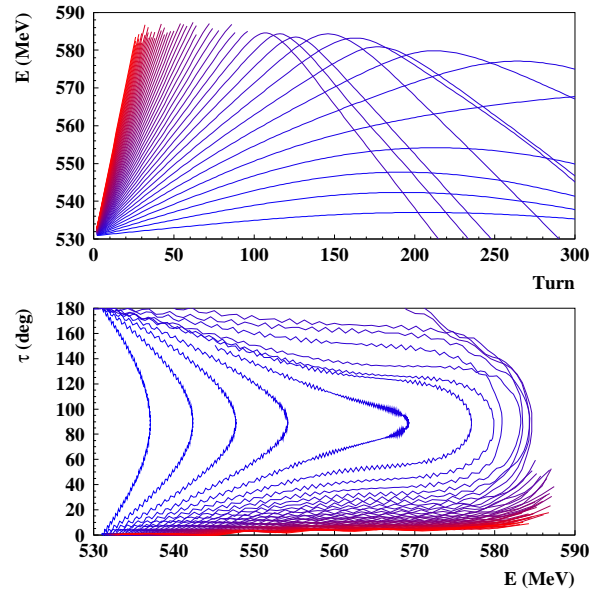}
}\caption[Tracking in the Ring Cyclotron]{
  Tracking results for escape extraction from the PSI Ring cyclotron.
  Top: Energy vs. turn number for orbits starting at $530\,\rm{MeV}$
  in the PSI Ring cyclotron. The rf-voltage increases from blue to red.
  Bottom: Phase $\tau$ of the rf at $\theta=0^\circ$ versus energy.
\label{fig_rfscan}}
\end{figure}

Fig.~\ref{fig_rfscan} shows results of direct orbit tracking\footnote{
  We use CYBORC (``Cyclotron Beam Orbit Calculator'') for the tracking,
  a ``C''-implementation of a 4th-order Runge-Kutta solver for the 
  equations of motion on a polar grid~\cite{Gordon84}.
  The field map of the PSI Ring~\cite{Seidel2021} that we used for the 
  tracking is based on measured data~\cite{Willax1969,Willax72,Adam1975} and 
  has been used in several other recent publications, for instance 
  Refs.~\cite{Yang2010,Bi2011,Frey2019}.
}, starting at $530\,\rm{MeV}$ and zero phase, which show that beam escape 
is possible for less than about $90$ turns, i.e. for an energy gain per 
turn of more than $(590\,\rm{MeV}-530\,\rm{MeV})/90\approx 670\,\rm{keV/turn}$.
This is indeed by a factor of about $4$ below the currently used 
acceleration voltage. 

Therefore the acceleration of the beam in the PSI Ring machine is so fast
that the beam approaches the bending limit before the beam phase is shifted
to $90^ \circ$. The beam would escape the magnetic field without any 
extraction device.
Hence the acceleration of the beam in the PSI Ring machine is so fast 
that it would reach the bending limit before the phase is shifted
to $90^ \circ$. The beam would escape the magnetic field without any 
extraction device.

Nonetheless the beam has to pass the fringe field region, where the 
negative $k$-values lead to a strong radial defocusing and vertical focusing. 
Both effects have to be compensated if one aims to make practical use of 
escape extraction. Furthermore, since the beam has to pass the $\nu_r=1$-resonance,
it is required to precisely control the first harmonic content of the field prior to
extraction.

\section{Extraction by Acceleration from the PSI Ring cyclotron}

\begin{figure}[t]
\parbox{8cm}{
\includegraphics[width=80mm]{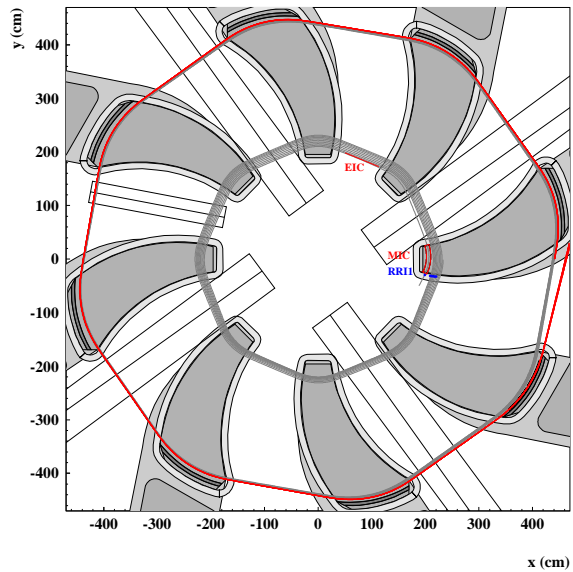}
}\caption[Turns in the Ring Cyclotron]{
    The iron shape of the PSI Ring cyclotron. Some turns of a (centered) 
    accelerated orbit are plotted on top. The first few turns after
    injection and the last few turns before extraction are shown in gray.
    The extracted orbits are shown in red.
    The last turn escapes the magnetic field at $\theta\ge 320^\circ$.
    The five rectangular boxes indicate the positions and size of the cavities,
    i.e. four main cavities and one flat-top cavity.
\label{fig_ring}}
\end{figure}

\begin{figure}[t]
\parbox{8cm}{
\includegraphics[width=80mm]{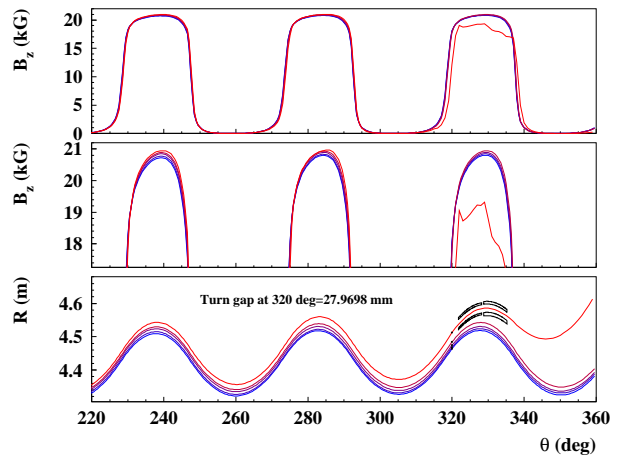}
}\caption[Extraction from Ring Cyclotron]{
    Top: Magnetic field along the last five turns (from blue to red).
    Center: Same as top, but zoomed in for details. Bottom:
    Radius versus azimuthal angle of the last five turns (blue to red).
    The last turn escapes the magnetic field at $\theta\ge 320^\circ$.
    The turn separation between the last two turns at $\theta=320^\circ$
    is more than $25\,\rm{mm}$ for an almost centered beam -- compared
    to $6\,\rm{mm}$ at $E=590\,\rm{MeV}$~\cite{hipa_acc}. The position of
    the main iron bars of a passive gradient corrector are shown as well.
\label{fig_xring}}
\end{figure}

Fig.~\ref{fig_ring} shows the iron shape of the PSI Ring cyclotron, some accelerated 
orbits and the last orbit which escapes the field without electrostatic extractor. 
The positions of the four accelerating cavities (plus one
flat-topping cavity) are indicated by the five rectangular boxes. The cavities 
provide enough energy gain per turn to extract the beam after about $185$
turns~\cite{hipa_acc}. Currently the Ring cyclotron uses an electrostatic
extractor to extract the beam at $E\approx 590\,\rm{MeV}$, i.e. before the
maximum field is reached. 
\begin{figure}[t]
\parbox{8cm}{
\includegraphics[width=80mm]{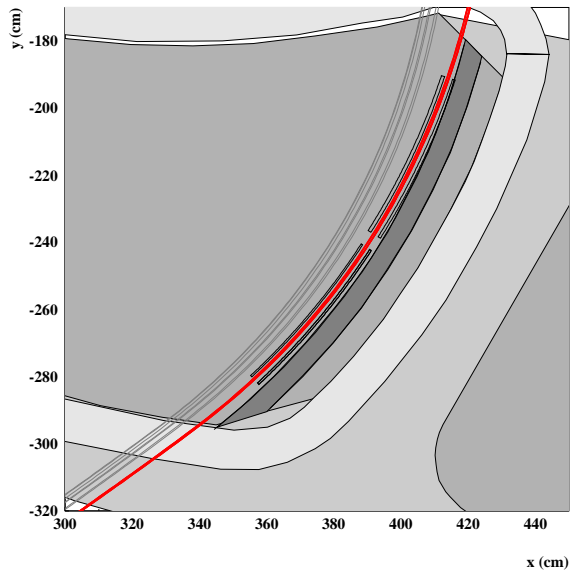}
}\caption[Last turns in the Ring Cyclotron]{Zoom of Fig.~\ref{fig_ring}.
    The thin black lines indicate the positions of the iron bars of 
    possible gradient correctors. They are (partially) in conflict with the 
    non-magnetic mechanical support welded to the poles (shown in 
    dark gray). Hence the Ring cyclotron could have been designed for
    escape extraction, but an upgrade is technically risky and difficult.
\label{fig_ringxx}}
\end{figure}
Figs.~\ref{fig_ring} and~\ref{fig_ringxx} show the escape extraction of nine 
orbits. Besides a central (``reference'') orbit, we tracked orbits with a 
different starting radius ($\pm 3\,\rm{mm}$), orbits with different initial 
energy ($\pm 0.1\,\rm{MeV}$), initial radial momentum ($\pm 5\,\rm{mm}$)
\footnote{To obtain the momentum in SI units, one has to multiply by $mc/R_\infty$.}
and rf phase ($\pm 3^\circ$). It is not only that the orbit escapes 
the field, as shown in Fig.~\ref{fig_xring}, but furthermore the field gradient 
is positive or zero up to the last sector before escape. 
Hence the orbit ``sees'' a substantial negative field gradient only for 
a short time. Hence a single gradient corrector might be sufficient to keep 
the beam radially compact. The magnetic field gradient is of the order of 
$-2\,\rm{kG/cm}$, i.e. a value for which a compensation might be possible,
at least in principle.

However, the upper and lower half-poles of the PSI Ring cyclotron are 
connected by a non-magnetic support at the (inner and) outer pole radius. 
The orbit of the extracted beam would come close to the support structure,
which is welded to the poles as a part of the vacuum chamber
(shown in dark gray in Fig.~\ref{fig_ringxx}).
This leaves few space for the installation of corrector magnets.
A modification of the support structure would imply severe practical 
difficulties and a long interruption of user data taking.
However in case of new cyclotron projects, the sketched 
extraction method seems to be a promising possibility. 

\section{Discussion and Outlook}

In new high intensity cyclotron project the design and shape of the 
magnetic fringe field and of mechanical components around it could be 
optimized for escape extraction. 
Since electrostatic extractors are notorious for high voltage breakdowns,
this option might specifically be interesting in cases where the number 
of acceptable beam interruptions is rigorously limited (as for instance in 
case of ADS). 
The negative field gradient in the fringe field requires the use of 
magnetic gradient correctors~\cite{Kleeven2001}. As shown in 
Fig.~\ref{fig_xring}, the 
field of the last turn is larger than that of the previous turn up to the
last two sectors and is significantly lower only in the last sector.
The extraction is therefore reasonably fast. A gradient corrector
magnet, no matter if active or passive, will almost certainly lead to
a lower number of beam interruptions than electrostatic extractors.
Due to Joho's ``$N^3$ law'' one can reasonably assume that any 
high-intensity and low-loss proton machine aiming for beam power 
in the multi-MW-range will require a high energy gain per turn.
The possibility of escape extraction might, under these conditions,
be regarded as a side-effect of high intensity operation.

The voltage and power that can be provided by a single cavity has 
practical limitations. The PSI main cavities, for instance, are
specified for power losses of $\le 500\,\rm{kW}$~\cite{Fitze}.
Hence the four main cavities of the PSI Ring cyclotron restrict the
maximum power loss to $\approx 2\,\rm{MW}$. If one takes this
as the state-of-the-art, then a multi-MW-cyclotron would require 
substantially more cavities than the PSI Ring cyclotron.
Furthermore the power loss in rf cavities is proportional to the
square of the cavity voltage~\cite{Grilli2016,hipa_acc}. Therefore
the cavity wall losses increase by a factor of four when the voltage
is doubled, but only by a factor of two when the number of cavities is
doubled. Hence the use of a higher number of cavities is beneficial for
the energy efficiency as well. A high number of cavities naturally suggests a
high number of sectors for high power proton cyclotrons. This requires 
space and hence an increase in radius. Though size seems to be an important
criterion, the size of a high intensity cyclotrons is negligible when compared
to the size of linacs for similar energies. The MYRRHA linac, designed
to provide $4\,\rm{mA}$ at $600\,\rm{MeV}$, for instance, has a length of
$400\,\rm{m}$~\cite{DEBRUYN2015}.

In the PSI machine, the effective turn separation between the last two turns
is, for a centered beam, $6\,\rm{mm}$, which can be enhanced up to $18\,\rm{mm}$ 
by betatron oscillations~\cite{hipa_acc}. This is still substantially 
smaller than the pole gap. Therefore the radial turn separation is as yet 
the bottleneck for high intensity operation.
Hence new high power cyclotrons aiming for power levels in the MW-regime, will 
likely be designed with a considerably larger extraction radius than the 
PSI Ring cyclotron, but not necessarily with a much larger pole gap.
Then the ratio of pole gap to extraction radius (and hence the required voltage
for escape extraction) will naturally be lower than (or equal to) the ratio of the
PSI Ring cyclotron and this will facilitate escape extraction even further. 
Due to these arguments we believe that it is worthwhile to further investigate
the feasibility of extraction purely by acceleration -- specifically in
high power cyclotrons. 

\section{Acknowledgements}

The figures have been generated with the CERNLIB (PAW) and XFig.

\bibliography{hix_paper}{}
\bibliographystyle{unsrt}

\end{document}